\documentclass{raa}

\usepackage{hyperref}
\usepackage{graphicx,times}             %for PS/EPS graphics inclusion, new
\usepackage{subfigure}  %% subfigure packet for figure combine
\usepackage{amsmath}  %%引用数学工具
\usepackage{caption} %%用于caption 居中

\graphicspath{{figures/}}
\begin{document}

   \title{IPS Observation  System  for Miyun 50m Radio Telescope and Its Acceptance Observation
%\,$^*$
%\footnotetext{$*$ Supported by the National Natural Science Foundation of China.}
}
%   \subtitle{I. Place Your Subtitle Here}

   \volnopage{Vol.0 (200x) No.0, 000--000}      %%preserved for Editor. DOn't remove!
   \setcounter{page}{1}          %%starting page, preserved for Editor. DOn't remove!

   \author{Xin-Ying Zhu
      \inst{1}
   \and Xi-Zhen Zhang
      \inst{1}
   \and Hong-Bo Zhang
      \inst{1}
   \and De-Qing Kong
      \inst{1}
   \and Hui-Peng Qu
      \inst{2}
   }
%% Here is an example of three authors come from different institutes.
%% For single author or all the authors from an institute, use "\inst{}" only

   \institute{National Astronomical Observatories, Chinese Academy of Sciences,
             Beijing 100012; {\it zhuxy@bao.ac.cn}\\
   \and
            Beijing Institution of Tracking and Telecommunication Technology ,Beijing 100094;\\
   }
%% Please give the E-mail address of the author, to whom future correspondence and
%% offprint requests will be sent.

   \date{Received~~2009 month day; accepted~~2009~~month day}

\abstract{ Ground-based observation of Interplanetary Scintillation(IPS) is an important approach of monitoring solar wind. A ground-based  IPS observation system is newly implemented on 50m radio telescope, Miyun station, National Astronomical Observatories, Chinese Academy of Sciences(NAOC). This observation system is constructed for purpose of observing the solar wind speed and scintillation index by using the normalized cross-spectrum of simultaneous dual-frequency IPS measurement. The system consists of a universal dual-frequency front-end and a dual-channel multi-function back-end specially designed for IPS. After careful calibration and testing , IPS observations on source 3C273B and 3C279 are successfully carried out. The preliminary observation results show that this newly developed observation system is capable of doing IPS observation.The system sensitivity for IPS observation can reach over 0.3Jy in terms of IPS polarization correlator with 4MHz bandwidth and 2s integration time.
}

 %%author_head in even pages
    \authorrunning{X.Y. Zhu, X.Z. Zhang, H.B. Zhang, D.Q. Kong \& H.P. Qu}
   \titlerunning{IPS Observation System for Miyun 50m Radio Telescope and Its Acceptance Observation }  % title_head in odd pages

   \maketitle
%% The author head (on even pages) and the title head (on odd pages) will be
%% automatically extracted from \author{} and \title{}. Whenever the title is too long,
%% you will be asked to supply a shorter one by inserting either \authorrunning{} or
%% \titlerunning{} before \maketitle. Anyway, you can specify your own heads.
%%
%%
%% Note: In the following text body of your manuscript, please note several differences from
%%       other major journals:
%% (1) \subsection{Please Capitalize the First Letter of Each Notional Word in Subsection Title}
%% (2) Please Capitalize the First Letter of Each Notional Word in all tables' captions

%
%________________________________________________ sections below
%

\section{INTRODUCTION}           %% first-level sections will be auto-capitalized
\label{sect:intro}

Interplanetary scintillation (IPS) is the random fluctuation in the intensity and phase
of electromagnetic waves passing through the interplanetary space. The fluctuation is
caused by refraction and deflection from the inhomogeneous plasma (solar wind) in
the interplanetary space(Zhang.~\cite{Zhang2007} ). Studying the solar wind impacts not only solar physics, space physics and geophysics, but also the related fields, such as aerospace activities, space communications, safety of humanity, and so on.

IPS observations with ground-based antennas have led to velocity estimates of the solar wind and also the structure estimates of distant compact radio sources(Hewish et al.~\cite{Hewish1969}; Armstrong et al.~\cite{Armstrong1972}). Such measurements, though indirect, give information out of the ecliptic plane and close to sun, where direct spacecraft have not yet been made(Scott et al.~\cite{Scott1983}). Several important IPS stations are using ground-based telescopes to do IPS observation, such as Cambridge (UK) (Hewish
et al.~\cite{Hewish1964}; Purvis et al.~\cite{Purvis1987}), Ooty (India) (Swarup et al.~\cite{Swarup1971}; Manoharan et al.~\cite{Manoharan1990}), Puschino (Russia)(Vitkevich et al.~\cite{Vitkevich1976}), STEL (Japan) (Asai et al.~\cite{Asai1995}),MSRT (China)(Zhang et al.~\cite{Zhang2001}; Wu et al.~\cite{Wu2001}). Many useful results are produced by such observations.

In 2005, A mega-project of science research on space weather monitoring, namely the Meridian Space Weather Monitoring Project (Meridian Project for short), has been approved by the Chinese government(Wang et al.~\cite{Wang2009}). As a group member of the Meridian Project, a ground-based IPS observation facility is sponsored and constructed with the objectives of observing solar wind speed and scintillation index.

Two methods could be adopted to observe IPS using single telescope. The first method is the single station single frequency observations (SSSF) (Hewish et al.~\cite{Hewish1964}; Manoharan et al.~\cite{Manoharan1990}; Liu et al.~\cite{Liu2009}). The second method is the single station dual frequency method (SSDF) (Scott et al.~\cite{Scott1983}). According to preliminary studies(Zhang et al.~\cite{Zhang2007}), compared to the SSSF method, the SSDF technique has the following
advantages: (1)Higher accuracy in the calculation of the characteristic frequency; (2)Small effects from
the variation of the solar wind parameters; (3)Higher sensitivity. The only additional cost for the SSDF
technique onto 50m radio telescope is for the receiver development. Thus we made efforts to build an IPS observation system with a dual-frequency front-end and dual-channel multi-function back-end to implement SSDF technology.

In the present paper, we outline the IPS observation system in section 2. IPS observations and results are given in section 3. Finally, a preliminary conclusion on IPS observation system is presented in Section 4.

%% Authors can give a citation as 'Michel et al. 1992'.
%% You may also use \cite, \citep and \citet for citation, and use Table~1 or Figure~1
%% and so forth. Using \ref and \label for cross-references of Tables/Figures
%% is a good way in adjusting/adding/removing text, tables or figures.

\section{OBSERVATION SYSTEM}
\label{sect:Obs}

  \begin{figure}
   \centering
   \includegraphics[width=0.5\textwidth, angle=270]{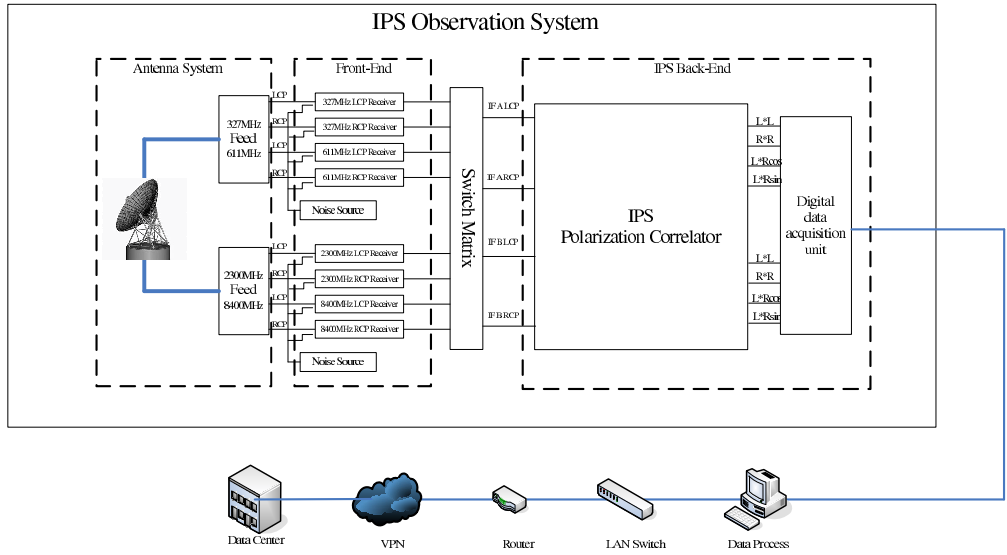}
   \caption{The block diagram of IPS observation system }
   \label{fig:i_sys}
  \end{figure}

 \autoref{fig:i_sys} illustrates the general layout of the IPS observation system as it relates to signal flow. IPS observation system consists of telescope system, front-end and back-end.
When doing IPS observation, the telescope is driven to aim at the target radio source. The radio wave emitted by the radio source impinges the telescope and is then focused and filtered by the selected working feed. The bandwidth limited radio frequency(RF) signal then be amplified and down-converted to an intermediate frequency(IF) signal by the front-end. The IF signal effectively lowers the transmission losses as the signal pass through the coaxial cables. The switch matrix is used to adjust the input and output at one's option. Each two IF signals represented two polarization of one working frequency are connected to one IPS polarization correlator, after power detection and correlation four components of input IF signals(L*L,R*R,L*Rcos,L*Rsin) are produced. Output components of IPS polarization correlator are then converted from an analog signal to a digital signal and recorded by digital data acquisition unit as original observation data. After processed by data process computer, both original observation data and preliminary scientific parameters such as solar wind speed and scintillation index are transferred to the data center through network.The system sensitivity for IPS observation can reach over 0.3Jy in terms of IPS polarization correlator with 4MHz bandwidth and 2s integration time.

\subsection{TELESCOPE SYSTEM}

\begin{figure}
   \centering
   \includegraphics[width=0.5\textwidth, angle=270]{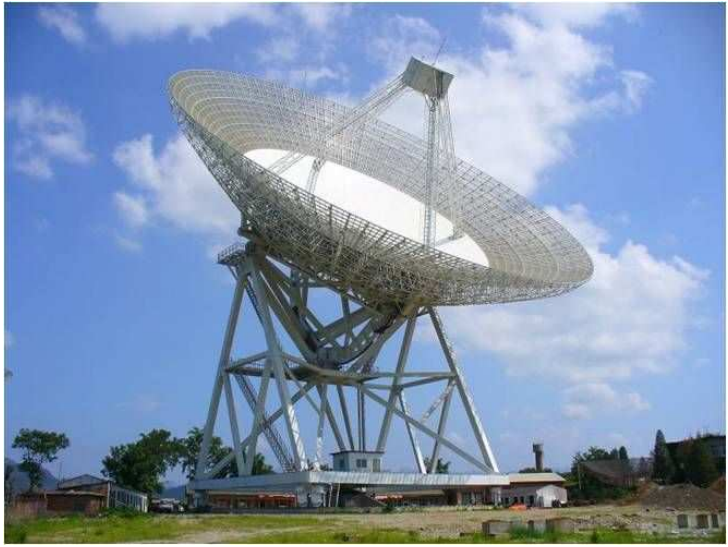}
   \caption{50m telescope in Miyun station }
   \label{fig:50m}
\end{figure}

The 50m Radio Telescope, which is now the biggest radio telescope in China, was built in 2006 for the project of Chinese Lunar Exploration. The telescope is located about 1.5 km south of the village of Bulaotun, a north part of the county of Miyun. In the vicinity of the telescope is the Miyun reservoir, which is the most important water source of capital, with a designed water storage capacity of 4.3 billion cubic meters. The telescope and the reservoir are surrounded by mountains, which provide a good electromagnetic environment for the telescope by preventing the RFI from outside.

The 50m radio telescope is operated by the NAOC.
Like most of the other large radio telescopes in the world, it has a azimuth-zenith mounting. The whole Telescope is mounted on wheels on a circular track, and the dish moves about a horizontal axis between the two A-like beam. The parabolic reflector is made from a solid filled panel in the inner 30m region and wire mesh in the outer part. All receivers and feeds are located at the prime focus. The 50m radio telescope is showed in \autoref{fig:50m} and the technical data of telescope are listed in \autoref{tab:50m} (Zhang et al.~\cite{Zhang2009}).

%
%               one-column-spanning table
%________________________________________ Table 2: Use_of_the routines
\begin{table}

\caption{ Technical Data of the Telescope}
\label{tab:50m}

%%Please Capitalize the First Letter of Each Notional Word in table's caption
\begin{center}
\begin{tabular}{ll}

\hline\noalign{\smallskip}
Reflector Diameter & $50 m$   \\
Aperture	& $1099m^2(S)$, $852m^2(X)$, $1119m^2(327MHz)$, $1119m^2(611MHz)$       \\
Pointing Accuracy &	$19"$\\
Azimuth Range &-270$^\circ ~\sim~+270^\circ $\\
Maximum Rotation Speed	& $60^\circ/min$\\
Elevation Range	& $from ~7^\circ ~to ~88^\circ$\\
Maximum Tilt Speed	&$ 30^\circ/min$\\
Total Weight	& $640 t$\\
Optics Prime focus & $F/D=0.35$\\
\noalign{\smallskip}\hline
\end{tabular}
\end{center}
\end{table}

\subsection{FRONT-END}
\begin{figure}
   \centering
   \includegraphics[width=0.5\textwidth, angle=270]{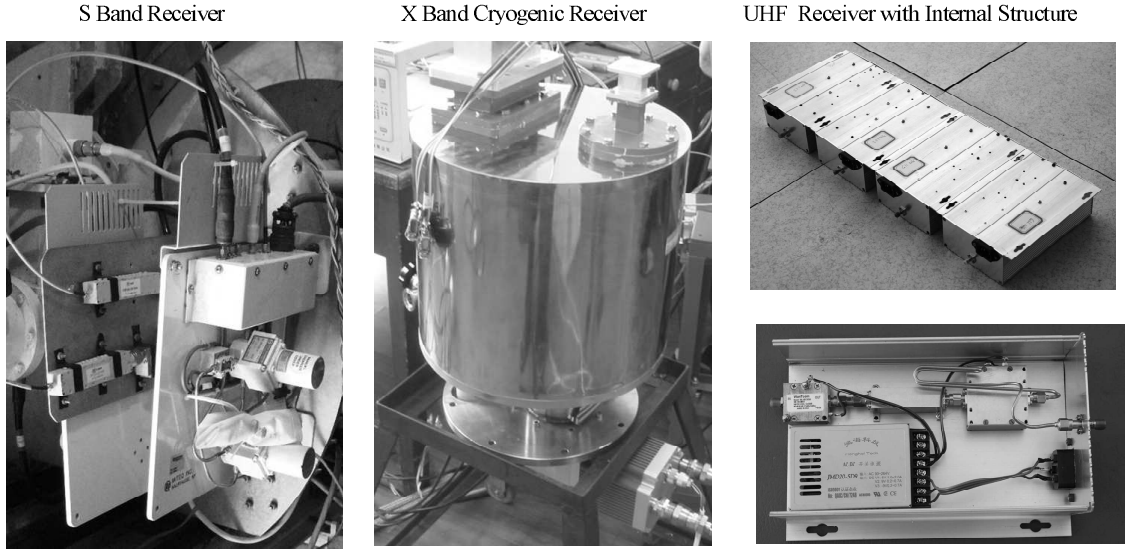}
   \caption{Front-end of IPS observation system. }
   \label{fig:f_e_all}
   \end{figure}

According to the design of IPS observation system, IPS front-end requires a total of 8 sets receivers, four sets for S/X dual-band feed and another four sets for UHF dual-band feed. Up to four of the 8 sets receivers could be selected to participate observation at the same time due to different need of science. Because of financial constraint, 3 existing receivers are used for S/X band, two room temperature receivers for S band and one cryogenic receiver for X band. Four new UHF receivers have already been developed, and will be installed onto the telescope after the Chang-E II mission next year. All available receivers for IPS front-end are showed in \autoref{fig:f_e_all} and the specifications of each receiver are listed in \autoref{tab:f_e}.

%
%               one-column-spanning table
%________________________________________ Table 2: Use_of_the routines
\begin{table}

\caption{ The Main Specifications of the Front-End}
\label{tab:f_e}

%%Please Capitalize the First Letter of Each Notional Word in table's caption
\begin{center}
\begin{tabular}{l l l l l l}

\hline\noalign{\smallskip}
No.&Item&S band&X band &UHF(327MHz) &UHF(611MHz)\\
1&Frequency Range&	2150-2450 MHz	& 8200-9000 MHz& 	307-347 MHz&	591-631 MHz \\
2&	Gain	& $\geq$ 65dB&	$\geq$ 65dB	&$\geq$ 65dB	&$\geq$ 65dB \\
3&	Receiver Noise Temperature&	$\leq$ 50K	&  $\leq $20K	& $\leq$ 55K	& $\leq$ 55K\\
4&	IF&	550-850 MHz &	100-900 MHz &	307-347 MHz 	&591-631 MHz  \\
5&	Polarization&	LCP\&RCP &	RCP	& X\&Y & X\&Y \\

\noalign{\smallskip}\hline
\end{tabular}
\end{center}
\end{table}

\subsection{BACK-END}
\begin{figure}
   \centering
   \includegraphics[width=0.5\textwidth, angle=270]{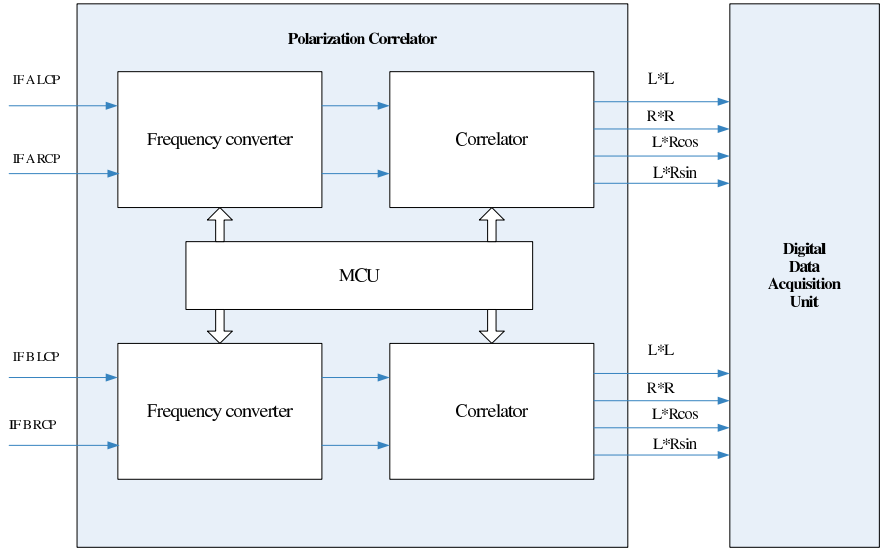}
   \caption{The block diagram of IPS back-end }
   \label{fig:b_e_all}
   \end{figure}

 A dual-channel multi-function back-end is designed exclusively for IPS observation. The back-end consists of a polarization correlator and a digital data acquisition unit, the block diagram of back-end is showed in \autoref{fig:b_e_all}. The input signal to polarization correlator is down-converted and amplified to make sure the mixer works in best condition. In order to prevent the RFI, the local oscillator frequency of the mixer and signal bandwidth could be chosen according to observation's actual need. In addition,the integration time could also be adjustable between 10ms to 2s with step size 10ms. Four output components ((L*L,R*R,L*Rcos,L*Rsin)) are produced after the correlation of each two input signals(L,R). A virtual instrument technology which consists of an workstation equipped with powerful application software and a plug-in sample board is used to perform the functions of digital data acquisition unit. The analog signal output by polarization correlation is captured by the plug-in sample board through its IO port, then captured signal has been converted from an analog signal to digital signal by A/D converter. After AD conversion, the signal is stored in FIFO Buffer to form a data block. The workstation equipped with this plug-in sample board reads the data blocks via the USB bus on a regular intervals, then makes some necessary calculations and saves the data according to the user-defined format.

\section{OBSERVATIONS AND RESULTS}
\label{sect:data}
In September 2011, acceptance observations of new IPS observation system were conducted at Miyun station after a period of system testing, debugging, and optimization. The IPS observation system intended to monitor the solar wind speed and scintillation index is capable of working at SSSF mode and SSDF mode. During SSSF mode, one of 327MHz,611MHz,2300MHz,8400MHz could be selected as the working frequency. During SSDF mode, both 327MHz/611MHz frequency group and 2300MHz/8400MHz frequency group could be selected to conduct the observation respectively. Since the scintillation decreases sharply in high frequency, strong source such as 3C273B, 3C279 are chosen for the acceptance observation. The radiation spectrum of 3C273B, 3C279 is flat, which means the source's flux density in 2300MHz is almost the same as value in 8400MHz. The key parameters of the acceptance observation with the new IPS observation system are listed in \autoref{tab:o_p}.

\begin{table}

\caption{ Key Parameters of acceptance Observation with the New IPS Observation System}
\label{tab:o_p}

%%Please Capitalize the First Letter of Each Notional Word in table's caption
\begin{center}
\begin{tabular}{l l }

\hline\noalign{\smallskip}

Observation Date  &  2011/09/20$\sim$2011/09/29,  2011/10/12$\sim$2011/10/20   \\
telescope 	& Miyun 50m Radio Telescope       \\
Source/Flux &	3C273B/41(Jy), 3C279/12(Jy) \\
Frequency/Bandwidth  &2300MHz/80MHz, 8400MHz/80MHz\\
Minimum Distance from Sun  &	3C273B/18$R_s$, 3C279/15$R_s$  ($R_s$: Sun Radius) \\
Sample Time	& Hardware 1ms, Software 10ms\\
Shortest Integrated Time& 240s\\

\noalign{\smallskip}\hline
\end{tabular}
\end{center}
\end{table}

\begin{figure}
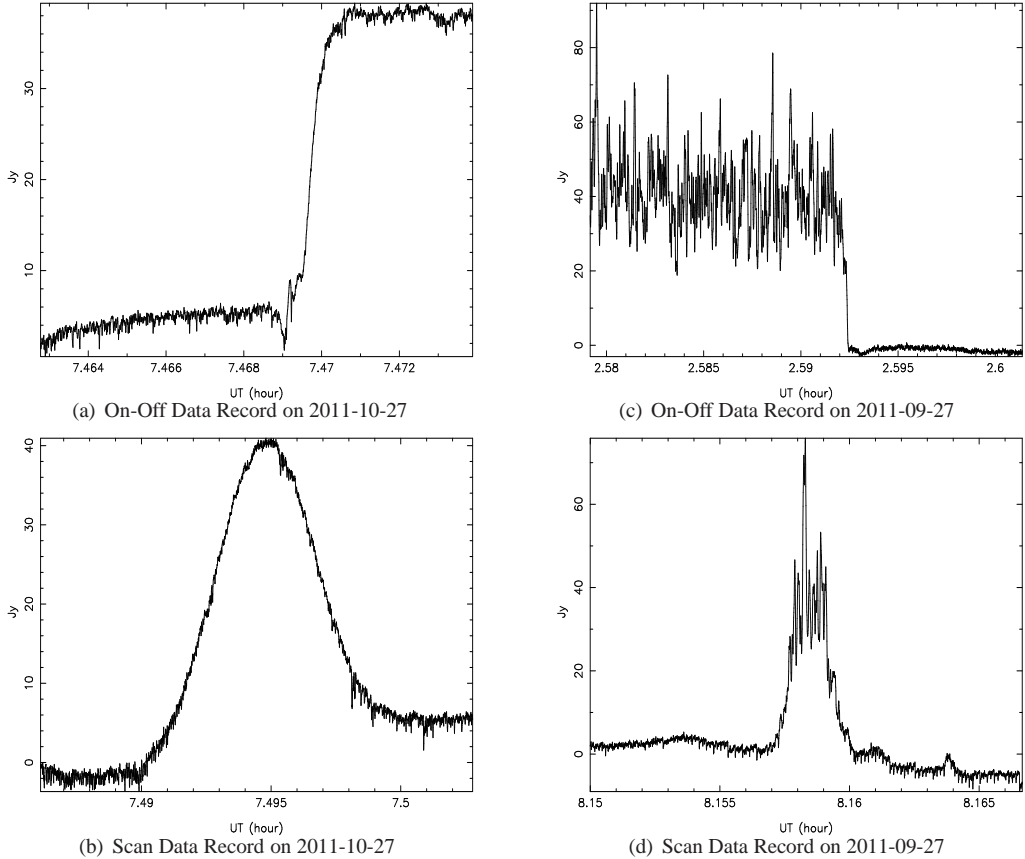

  \begin{minipage}{0.5\linewidth}
    \subfigure[On-Off Data Record on 2011-10-27]{   \includegraphics[width=53mm,height=62mm, angle=270]{ms1009fig5-1.ps}  }
    \subfigure[Scan Data Record on 2011-10-27]{   \includegraphics[width=53mm,height=62mm, angle=270]{ms1009fig5-2.ps}}
   \end{minipage}
  \begin{minipage}{0.5\textwidth}
    \subfigure[On-Off Data Record on 2011-09-27]{   \includegraphics[width=53mm,height=62mm, angle=270]{ms1009fig5-3.ps}}
    \subfigure[Scan Data Record on 2011-09-27]{   \includegraphics[width=53mm,height=62mm, angle=270]{ms1009fig5-4.ps}}
  \end{minipage}
  \caption{(a)$\&$(b)Raw observation data from 3C273B on October 27,2011.(c)$\&$(d):Raw observation data from 3C273B on September 27,2011.The ordinate on behalf of power is scaled linearly(2Jy per division) and the abscissa on behalf of UTC is scaled linearly(0.001hour per division).}
   \label{fig:r_d}
\end{figure}

When radio wave emitted from distance source pass through interplanetary space, the intensity and phase of the radio wave will scintillate because of the existence of solar wind. The scintillation which carries a variety of physical information on the spatial distribution of space plasma is proportional to the negative fourth power of the heliocentric distance. The distance where the weak scintillation regime sets in is a function of the frequency, See \autoref{tab:m_d_f}. From \autoref{tab:m_d_f}, short wavelengths are necessary to observe closer to the sun and are much more difficult to observe compared with long wavelengths. For this reason, the ability to detect the changes of received radio signal on short wavelength is an important criteria to verify the whole IPS observation system. The S band raw observation data of the IPS observation system are showed in \autoref{fig:r_d}.

\begin{table}

\caption{ The minimum distance the weak scintillation condition begins to hold as
a function of frequency.}
\label{tab:m_d_f}

%%Please Capitalize the First Letter of Each Notional Word in table's caption
\begin{center}
\begin{tabular}{l l l l l l l l}

\hline\noalign{\smallskip}
Frequency(MHz) & 150& 327 &900 &2000 &5000 &7500& 20000\\
Heliocentric distance ($R$: Sun Radius) &60$R$  &35$R$ & 18$R$ & 10$R$ & 5$R$ & 4$R$ & 2$R$\\

\noalign{\smallskip}\hline
\end{tabular}
\end{center}
\end{table}

\begin{figure}
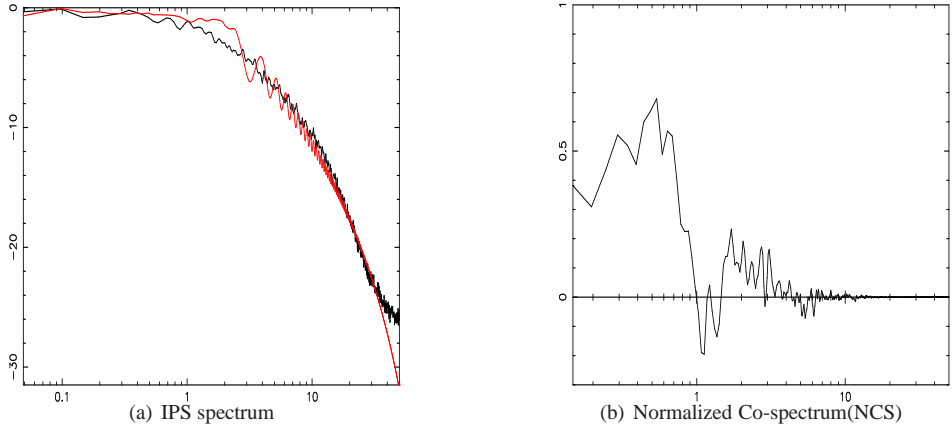

  \begin{minipage}{0.5\linewidth}
    \subfigure[IPS spectrum ]{   \includegraphics[width=53mm,height=52mm,angle=270]{ms1009fig6-1.ps} }
  \end{minipage}
  \begin{minipage}{0.5\textwidth}
    \subfigure[Normalized Co-spectrum(NCS)]{      \includegraphics[width=53mm,height=52mm,angle=270]{ms1009fig6-2.ps}}
    \end{minipage}
  \caption{(a):IPS spectrum of observation data from 3c237b on September 27,2011( Red line is the theoretical fitting curve).(b)Normalized Co-spectrum(NCS) between the pairs of radio frequencies 2300/8400 in MHz.Scales are logarithmic for frequency and power density (2dB per division) and linear for NCS.}
   \label{fig:result}

\end{figure}

\begin{figure}
   \centering
   \includegraphics[width=0.5\textwidth, angle=270]{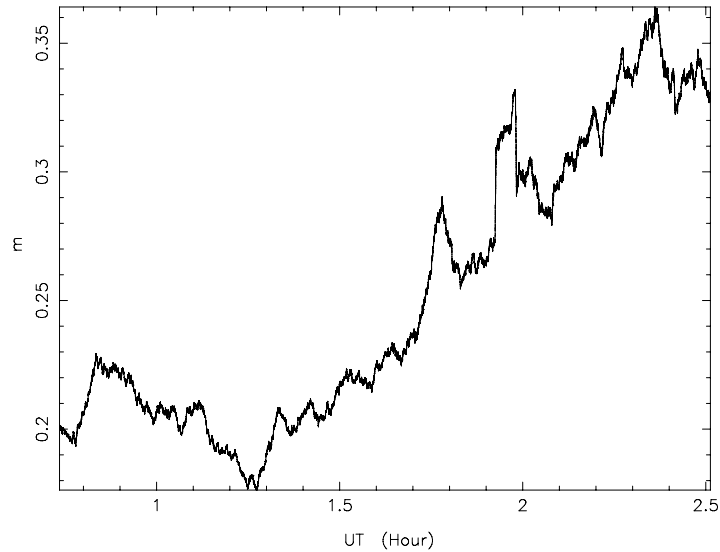}
   \caption{Scintillation index of observation data from 3c237b on September 27,2011 }
   \label{fig:index}
   \end{figure}

In \autoref{fig:r_d} (a), we show the OFF-ON raw data from 3C273B on October 27,2011 (OFF: The telescope keeps track of the target with a constant deviation; ON: The telescope keeps track of the target). At this time, the angular distance between the 3C273B and the sun is large and the observation data show that the OFF-ON noise fluctuation of received signal are almost the same and the signal received by the observation system is stable without RFI. In \autoref{fig:r_d} (c), the OFF-ON raw data from 3C273B on September 27 is presented. Because 3C273B is close to the sun at this time, big difference between OFF-ON noise fluctuation of received signal are observed. When the telescope targets OFF the 3C273B, the noise fluctuation of received signal is small just reflect the system's own noise fluctuation. When telescope targets ON the 3C273B, the severe fluctuation of the received signal strength shows that the radio signal emitted by 3C273B has been scintillated by solar wind. Comparison of \autoref{fig:r_d}(a) and \autoref{fig:r_d}(c) proves that the new developed IPS observation system have the ability to observe the interplanetary scintillation. \autoref{fig:r_d} (b) and \autoref{fig:r_d} (d) show raw data when telescope scan 3C273B on October 27 and September 27 respectively. The noise fluctuation of the whole telescope beam are almost the same in \autoref{fig:r_d} (b) while main lobe fluctuate severely in \autoref{fig:r_d} (d) because of the scintillation caused by solar wind.
The observation data from 3C273B on September 27, 2011 are used to do data analysis. After preliminary data processing, IPS spectrum and normalized cross-spectrum(NCS) of pairs of radio frequencies 2300MHz/8400MHz are obtained. The IPS power spectrum of 2300MHz is showed in \autoref{fig:result} (a), in which Red line represents the theoretical fitting curve. \autoref{fig:result} (b) shows the normalized cross-spectrum(NCS) of pairs of radio frequencies 2300MHz/8400MHz. According to theoretical fitting curve, the solar wind speed is about 630Km/s in the direction perpendicular to line of sight. The scintillation index of this observation is showed in \autoref{fig:index}.

\section{Conclusions}
\label{sect:conclusion}
According to preliminary result, the new developed IPS observation system is now capable of doing interplanetary scintillation observation.

\begin{acknowledgements}
This work was funded by the Meridian Space Weather Monitoring Project.
\end{acknowledgements}

\label{lastpage}

\end{document}